# A Comparative Review of Methods to Create a Composite Index for Sustainable and Inclusive Wellbeing

da Silva Vieira, Ricardo[a]*, Biggeri, Mario[b], Benczur, Peter[c], Costanza, Robert[d], Eastoe, Joseph[d], Hirvilammi, Tuuli[e], Kubiszewski, Ida[d], Mazziotta, Matteo[f], Mulder, Kenneth[g]; Muroya, Taketo[h], O'Connor, Kelsey J.[i], Sarracino, Francesco[i]; Rigas, Nikos[i], Giovannini, Enrico[j]; Hoekstra, Rutger[k]; Hopp, Daniel[l]; Horlings, Edwin[m]; Krylova, Petra[c]; Melchiorri, Michele[c]; Tapia, Heriberto[n]; Smallenbroek, Oscar[c]

[a] MARETEC, LARSYS, Técnico Lisboa, Universidade de Lisboa, Avenida Rovisco Pais 1, 1049-001 Lisboa, Portugal.

[b] Department of economics and Management, University of Florence, Via delle Pandette, 9, 50127, Firenze, Italy.

[c] Joint Research Centre, European Commission, Via E. Fermi, 2749, I-21027 Ispra (VA), Italy.

[d] Institute for Global Prosperity, University College London. 9-11 Endsleigh Gardens, London WC1H 0EH, UK.

[e] Tampere University, Kalevantie 4, 33100 Tampere, Finland.

[f] Italian Institute of Statistics, via C. Balbo 16, Rome, Italy.

[g] School of Natural Sciences, Hampshire College, 893 West St, Amherst, MA 01002, USA.

[h] Organisation for Economic Co-operation and Development, 2 rue André Pascal, 75016 Paris, France.

[i] STATEC Research, 12, boulevard du Jazz, L-4370, Belvaux, Luxembourg.

[j] University of Rome "Tor Vergata" and Italian Alliance for Sustainable Development (ASviS), Via Orazio Raimondo, 18, 00173 Roma, Italy

[k] Institute of Environmental Sciences, Leiden University, Van Steenis Building, Einsteinweg 2, 2333 CC Leiden, The Netherlands.

[l] United Nations Trade and Development, Palais des Nations 8-14, Av. de la Paix CH-1211 Geneva 10 Switzerland.

[m] CBS – Statistics Netherlands, P.O. Box 24500, 2490 HA The Hague, The Netherlands.

[n] United Nations Development Programme, One United Nations Plaza, New York, NY 10017 USA.

* Corresponding author ricardosilvavieira@tecnico.ulisboa.pt

## ABSTRACT

Societal goals need to shift from over-reliance on gross domestic product (GDP) to broader aspects of sustainable and inclusive wellbeing (SIW). However, defining SIW and eventually measuring it with a single number is problematic because it involves many subjective and objective contributors that combine in complex, non-linear ways. Conventional approaches either use linear weighted averages or reduce SIW to subjective wellbeing alone. Neither is sufficient.

This paper reviews aggregation methods for SIW against nine conditions derived from needs theory and strong sustainability: limited substitutability, penalisation of imbalances, non-linear transformations, respect for environmental ceilings, respect for lower limits, a formative measurement model, no correlation requirement, distributional sensitivity, cross-border spillovers, and intertemporal aggregation.

We compare 13 methods—from simple arithmetic means to penalty-based indices, outranking multicriteria, data envelopment analysis, and insights from ecology, neuroscience, and machine learning. Our illustrative example shows that aggregation choices change significantly country rankings. Compensatory methods create similar rankings. No single method satisfies all nine conditions.

We conclude that a future SIW composite indicator will require combining methods across levels: non-linear normalisation, non-compensatory aggregation, and measurement-level choices for inclusiveness and spillovers. This paper provides a step towards the headline aggregated indicator advocated by the UN High-Level Expert Group on Beyond GDP.

# 1 INTRODUCTION

Measuring sustainable and inclusive wellbeing (SIW) has moved from academic debate to urgent policy action (High-Level Expert Group on Beyond GDP, 2026).

GDP, the most familiar metric of national performance, measures economic output but does not take the further step of measuring wellbeing itself. Moreover, it cannot tell us whether wellbeing is achieved within environmental limits or shared inclusively (Jansen et al., 2024). Recognising this gap, the United Nations Secretary-General appointed a High-Level Expert Group on Beyond GDP (HLEG). Its final report delivered a stark message: delay in adopting better measures of progress is no longer an option (High-Level Expert Group on Beyond GDP, 2026). The report called for measuring progress in terms of equitable, inclusive, and sustainable wellbeing, and proposed the creation of a scientific committee to further develop headline aggregate indicators that transparently and rigorously capture the key dimensions of progress. This later reflects the fact that the HLEG did not propose a composite index structure based on its dashboard. This omission is not a shortcoming—it reflects precisely the difficulties we address in this paper. How should one aggregate heterogeneous dimensions without imposing arbitrary weights or unrealistic substitutability between SIW domains?

We define SIW as a state of wellbeing that is sustainable, inclusive, and attentive to the future. It encompasses current wellbeing (multidimensional quality of life), inclusion (the distribution of wellbeing across people and places), and sustainability (the maintenance of conditions for future wellbeing). This conception draws on the Brundtland Commission (1987), the Stiglitz et al. (2009), and the UN High-Level Expert Group on Beyond GDP (2026). A large body of research shows that societal wellbeing depends on many determinants, spanning subjective domains (e.g., life satisfaction) and objective conditions (e.g., health, education, material security, social conditions, and environmental quality) (Kubiszewski et al., 2025), their distribution and future prospects. Importantly, these relationships are rarely linear. Many determinants display diminishing returns, thresholds, or optimum levels. They can be interdependent, and shortfalls in basic needs or environmental conditions cannot always be compensated by improvements elsewhere (Doyal and Gough, 1991; Martínez-Alier, 1995; Max-Neef, 1991). These features are central to SIW, where environmental constraints and social floors limit acceptable trade-offs.

A single composite index to measure SIW would offer clear advantages. It would help track progress over time, compare countries, communicate performance succinctly, and identify policy bottlenecks across SIW domains. Yet creating an index that is both theoretically sound and methodologically robust has proven difficult, hence the decision of the HLEG not to propose such headline measures (they did provide illustrative examples of a very different strategy, theory-based adjusted GDP type monetary measures).

Most used approaches to aggregating domains into overall indices—common in social sciences, psychology, and economics—typically use weighted or unweighted arithmetic averages (European Commission. Joint Research Centre., 2024). Adopting such a practice to wellbeing raises important questions, however. Averaging assumes full substitutability (e.g., income can offset environmental damage), linearity without thresholds, and rarely impose environmental or social limits. Other approaches such as the World Happiness Report (Helliwell et al., 2024), treat subjective wellbeing as a single overarching measure, allowing trade-offs between predictors while omitting environmental constraints central to Earth system science and ecological economics. These limitations point to an unresolved issue: how can we create a composite index of SIW that accommodates multidimensionality, non-linearities, and limited substitutability—without reducing SIW to a weighted sum or to subjective wellbeing alone?

This paper answers the HLEG's call by reviewing methodological tools and exploring what frameworks can be adapted to SIW measurement. We do not propose a final SIW composite index.

This paper contributes by comparing leading approaches, clarifying their assumptions, illustrating them with a concrete example, and proposing recommendations for a future SIW composite index. The outline is as follows. Section 2 defines SIW and its key properties. Section 3 reviews existing aggregation approaches and their limitations. Section 4 reviews methods primarily from statistics, economics, and decision science that go beyond averaging and full compensability. Section 5 draws insights from other scientific fields that face similar issues. Section 6 illustrates selected approaches with an example. Section 0 discusses implications and limitations. Section 8 proposes recommendations. Section 9 concludes.

# 2 SUSTAINABLE AND INCLUSIVE WELLBEING AND CONDITIONS FOR A COMPOSITE INDICATOR

SIW comprises three interlinked dimensions. Current wellbeing is multidimensional, encompassing material, relational, and mental domains that together shape quality of life at the individual and societal level (McGregor and Pouw, 2016). No single domain is sufficient on its own. Relevant determinants include health, education, employment, social relationships, income, housing, security, and environmental quality (Kubiszewski et al., 2025; OECD, 2024). Inclusion concerns how these determinants are distributed across individuals, groups, and places—including inequalities in income, wealth, gender, poverty, and environmental harms, as well as cross-border spillovers (Boskovic et al., 2026). Sustainability concerns the long-term maintenance of biophysical and social conditions that support wellbeing for current and future generations, including economic, human and social capital, climate stability, biodiversity, natural capital, and institutional capacity (Boskovic et al., 2026; Jansen et al., 2024).

The multidimensional character of SIW poses two fundamental challenges to standard linear aggregation. First, SIW dimensions and even domains are not fully compensatory. Need theories argue that core human needs are non-additive and only weakly substitutable (Doyal and Gough, 1991; Max-Neef, 1991). Deficits in the domains of health or social connection cannot be fully compensated by gains in income. Improvements in material conditions can even generate trade-offs by undermining relational or mental wellbeing (McGregor and Pouw, 2016). Strong sustainability reinforces this point: natural systems provide essential and often non-substitutable contributions to wellbeing at the dimension level (Benczur et al., 2025; Costanza et al., 1997). Environmental degradation cannot be offset by increases in production or economic capital. Staying within planetary boundaries is a necessary condition for SIW, one that cannot be substituted for with more income (Rockström et al., 2009; Steffen et al., 2015).

Second, SIW dimensions and domains exhibit thresholds, saturation effects, and limiting factor dynamics. Beyond certain levels, additional gains in income, consumption, or even equality produce diminishing or even (almost) zero marginal returns. Need satisfaction is therefore characterised by sufficiency rather than unlimited expansion. Overall wellbeing depends less on maximising a few domains than on maintaining sufficient levels across all essential domains (Costanza et al., 1997). Research on wellbeing and energy use similarly shows saturation relationships (Steinberger and Roberts, 2010). Wellbeing functions more like a constrained system than a linear sum of independent domains.

Taken together, these insights suggest that SIW should be conceptualised as a balanced configuration of interacting dimensions with thresholds, diminishing returns, and limited substitutability. This perspective further underlines the need for a formative measurement model, where indicators define the construct rather than reflect a single latent variable (Mazziotta and Pareto, 2019). In a formative model, dimensions are qualitatively different and not interchangeable; removing one indicator changes the meaning of the construct itself, and correlation patterns between indicators are less of a concern.

From these conceptual foundations, we derive nine conditions that any SIW composite indicator should satisfy (Table 1). These conditions serve as normative benchmarks against which aggregation methods can be assessed.

**Table 1. Conditions for a SIW composite indicator**

| # | **Condition** | **Meaning** |
|---|---|---|
| **(i)** | Formative measurement model | Indicators cause the construct, not vice versa |
| **-** | No correlation requirement | Dimensions need not be highly correlated |
| **(ii)** | Distributional sensitivity (inclusiveness) | Inequality within countries should be penalised, not just average performance |
| **(iii)** | Cross-border spillovers | One country's wellbeing should not be achieved by degrading another's |
| **(iv)** | Limited substitutability | No dimension can be fully compensated by gains in another |

| # | Condition | Meaning |
|---|---|---|
| **(v)** | Penalisation of imbalances | Uneven performance across dimensions should lower the overall score |
| **(vi)** | Intertemporal aggregation | Current wellbeing should not hamper the conditions for future wellbeing |
| **(vii)** | Non-linear transformations | Diminishing returns, saturation, and thresholds must be captured |
| **(viii)** | Respect for upper limits (environmental ceilings) | Planetary boundaries are constraints, not optimisable variables |
| **(ix)** | Respect for lower limits (minimum acceptable thresholds for human needs) | Minimum thresholds for human needs must be satisfied |

Conditions (iv) through (vii) are closely related. Limited substitutability (iv) sets the basic rule that deficits cannot be fully offset. Penalising imbalances (v) extends this by treating balanced profiles as intrinsically preferable, even when average performance is held constant. Non-linear transformations (vii) operationalise saturation and threshold effects.

Condition (ii)—inclusiveness—differs from imbalance across dimensions (v): it concerns distribution of outcomes within a dimension across individuals or groups, not across dimensions for a single country. Condition (vi)—intertemporal aggregation—stands apart: unlike the others, it concerns having measures of future wellbeing and how o aggregate these with current wellbeing. Condition (viii)—respect for upper limits—reflects the strong sustainability principle that environmental limits are not negotiable. Exceeding a planetary boundary should not be compensable by superior performance elsewhere. Condition (ix)—respect for lower limits—ensures minimum quality of life and needs satisfaction.

The nine conditions operate at different levels of composite indicator construction. Condition (iii) (cross-border spillovers) is primarily a matter of indicator measurement (e.g., consumption-based rather than territorial accounts). Conditions (ii) and (vii) can be addressed at the normalisation stage—through concave functions or by including standard deviations as penalty terms—but also partly at aggregation. Conditions (iv), (v), (vi), (viii), and (ix) are primarily aggregation concerns: how normalised indicators are combined into a final score.

Critically, some conditions can be addressed at multiple levels, and no single method needs to satisfy all nine conditions on its own. This opens the possibility of combining different choices across levels: for example, using consumption-based indicators (iii), non-linear normalisation (vii), and penalty-based aggregation (iv, v, viii, ix). We return to this modular perspective in the discussion and recommendations.

## 3 BASELINE AGGREGATION APPROACHES AND THEIR LIMITATIONS

Composite indicators involve choices at multiple levels: how indicators are selected, measured, normalised, and aggregated. This section focuses on the most widely used aggregation-level methods. They are blind to the conditions iv, v and vii-ix, and their practical implementation is strongly embedded in a reflective and not a formative logic (i). Methods addressing those gaps are discussed in Section 4.

All methods in this section operate at the aggregation level only, assuming indicators have already been measured and normalised.

## 3.1 Arithmetic mean (unweighted and weighted)

The arithmetic mean is the most common aggregation method. It sums normalised scores, either with equal or pre-defined weights. It is a special case of the generalised mean family (Eq. 1), obtained when parameter $p = 1$.

$$\boldsymbol{M_p}(\boldsymbol{x_1}, \dots, \boldsymbol{x_n}) = \left(\frac{\sum_{i=1}^{n} w_i x_i^p}{\sum_{i=1}^{n} w_i}\right)^{\frac{1}{p}} \qquad \text{(Eq. 1)}$$

where, $M_p$ is the function to be modelled (e.g., SIW), $x_i$ the domains, $w_i$ the weights and $p$ a parameter from -∞ to +∞.

The arithmetic mean has been used in indices such as the OECD Better Life Index, or the UK's Office of National Statistics' Life Satisfaction Index. The arithmetic mean aligns with a reflective measurement model (condition i). It assumes full substitutability (condition iv): a deficit in one dimension can be fully offset by a surplus elsewhere. It imposes no penalty for imbalances (condition v): a lopsided country can score as well as a balanced one at the same average. It is linear (condition vii): it cannot capture diminishing returns or thresholds. It treats environmental ceilings and lower limits (conditions viii, ix) as optimisable rather than hard constraints. It.

## 3.2 Geometric mean

The geometric mean is the $n$th root of the product of $n$ indicators (Eq. 1 with $p = 0$). It is partially non-compensatory: a very low score strongly depresses the overall score.

The United Nations' Human Development Index makes use of geometric mean. It aligns better with formative logic (i). It considers substitutability (iv) (a proportional deficit can be fully offset by another proportional gain) but penalises imbalances (v) to some extent. However, it remains log-linear (vii): it captures proportionate changes but not thresholds or saturation. It does not respect upper or lower limits (viii, ix) as hard constraints. A drawback: it requires strictly positive normalised scores, problematic when min-max normalisation yields zeros.

## 3.3 Weighted linear additive model (participatory MCDA)

The participatory weighted linear additive model (LAM hereafter) extends the arithmetic mean by eliciting weights through participatory processes (Banville et al., 1998; Munda, 2004):

$$\boldsymbol{LAM}(\boldsymbol{x_1}, \dots, \boldsymbol{x_n}) = \sum_{i}^{n} \boldsymbol{w_i x_i} \quad \text{(Eq. 2)}$$

where $x_i$ are the domains, $w_i$ the weights. LAM makes normative trade-offs explicit, enhancing legitimacy. However, it shares the same limitations as the arithmetic mean regarding substitutability (iv), imbalance penalisation (v), linearity (vii), and thresholds (viii, ix). It remains compensatory unless explicitly restricted (e.g., by veto thresholds or weight bounds).

## 3.4 Principal Component Analysis (PCA)

PCA is a data-driven dimension-reduction technique. It identifies linear combinations of correlated indicators that capture maximum variance. The first principal component (PC1) is sometimes used as a composite index.

PCA is fundamentally misaligned with a formative approach and the dimensions of SIW. It is a data reduction technique which assumes zero measurement error rather than a formative measurement model (i). It privileges highly correlated, high-variance indicators, underrepresenting essential but weakly correlated dimensions such as environmental quality. It can assign negative weights when trade-offs exist, producing a "development–environment trade-off" axis rather than an SIW-aligned index. Weights are unstable across samples and time (Mazziotta and Pareto, 2019). PCA does not address inclusiveness (ii), spillovers (iii), substitutability (iv), imbalances (v), non-linearity (vii), or thresholds (viii, ix). We include it not as a recommended method but as a diagnostic warning: it answers what varies most, not what matters most. In some special cases, however, it may yield meaningful domain-level interim aggregates, like in the case of the Social Progress Index (Harmacek, 2026).

# 4 METHODS THAT ADDRESS SIW CONDITIONS

The methods in Section 3 share common blind spots: linearity and full compensability. This section reviews methods that address these gaps. They operate at different levels: indicator measurement (consumption-based accounting), normalisation (non-linear transformations), and aggregation (penalty-based indices, outranking, DEA). Some span multiple levels.

## 4.1 Indicator measurement: consumption and income-based approaches

Standard indicators are measured territorially: $CO_2$ is counted where emitted, income where earned. This ignores cross-border spillovers: a country can appear sustainable by importing emissions-intensive goods. Consumption-based accounting reallocates emissions to the final consumer; income-based accounting reallocates to the income receiver (Domingos, 2015). Territorial approaches favour countries with resource-intensive imports; consumption-based approaches favour exporters of resource-intensive products; income-based approaches tend to penalise exporters of products with high use-stage impacts (e.g., oil). A combination of accounting methods may be most informative.

These operate at the measurement level, directly addressing cross-border spillovers (condition iii). Data availability (e.g., global input–output tables) is a limitation (Boskovic et al., 2026; High-Level Expert Group on Beyond GDP, 2026).

## 4.2 Outcome-based regression (regression on life satisfaction)

Outcome-based regression estimates the contribution of SIW determinants by regressing a wellbeing proxy (typically life satisfaction) on explanatory variables (Burger et al., 2026; Helliwell et al., 2024). Although the index here remains subjective wellbeing and the regressions are more to uncover its drivers, the regression coefficients could be interpreted as implicit weights. The approach captures non-linearities (e.g., log income) and diminishing returns (Easterlin, 2015; Kahneman and Deaton, 2010).

This could be seen as a weight derivation approach. It can capture non-linear transformations (condition vii) and is empirically grounded. However, it reduces SIW to subjective wellbeing alone, omitting objective and environmental conditions (Costanza et al., 2016; Kubiszewski et al., 2025). Regression weights reflect statistical associations, not normative importance. Additive models struggle with threshold effects and non-substitutability (condition iv). It is a useful empirical tool, more suited to the current wellbeing dimension alone, but insufficient as a standalone SIW aggregation method.

## 4.3 Normalisation: non-linear transformations for saturation and diminishing returns

Before aggregation, indicators can be transformed to reflect diminishing returns, saturation, or threshold effects. Common transformations include logarithmic (diminishing marginal gains), logistic (S-shaped curves with lower and upper asymptotes), and power functions (concave shapes). These transformations embed substantive assumptions about how each determinant contributes to SIW. Income, for example, is often log-transformed to reflect that an extra dollar matters more to a poor person than to a rich one (Steinberger and Roberts, 2010). The choice of transformation requires empirical or normative justification.

These operate at the normalisation level and directly address non-linear transformations (condition vii).

## 4.4 Penalty-based indices: AMPI and MSI

### 4.4.1 Adjusted Mazziotta-Pareto Index (AMPI)

AMPI penalises imbalance across dimensions (Mazziotta and Pareto, 2016). It normalises using fixed goalposts (enabling temporal comparability), then calculates:

$$AMPI_i = M_{r_i} \pm (S_{r_i} * cv_i) \text{ , (Eq. 3)}$$

where $Mr_i$ is the mean, $Sr_i$ the standard deviation, and $cv_i$ the coefficient of variation for the indicator $i$. The penalty term ($Sr_i \cdot cv_i$) penalises unbalanced profiles.

AMPI operates at normalisation and aggregation (hence, out of scope for conditions ii, iii, vi). It addresses penalisation of imbalances (v) and is partially non-compensatory (iv). It respects formative logic (i) and enables temporal comparability. It addresses non-linearity (vii) but does not address upper/lower limits (viii, ix).

### 4.4.2 The Multidimensional Synthesis Indicator (MSI)

MSI allows substitutability to vary with achievement level (Biggeri and Mauro, 2018; Mauro et al., 2018):

$$MSI_i = 1 - [\frac{1}{k}\sum_{j=1}^{k}(1 - x_{ij})^{g(x_i)}]^{\frac{1}{g(x_i)}} \text{ , (Eq. 4)}$$

where $k$ is the number of dimensions, and $g(x_i)$ is a real-valued function of the unit's achievements. $g(x_i) \geq 1$ governs substitutability. Lower achievement restricts compensation (higher $g$); higher achievement allows more compensation ($g \to 1$). This implements dynamic substitution: poorer countries cannot trade off dimensions as freely as richer ones. However, fit penalises fewer poor outcomes than the geometric mean and for lower values does not collapse to zero as the geometric mean.

MSI operates at aggregation (hence, out of scope for conditions ii, iii, vi). It addresses limited substitutability (iv) and penalisation of imbalances (v) and enables temporal comparability (vi). It aligns with formative logic (i). It is flexible and could include upper/lower limits (viii, ix), it does not address nonlinearity (vii).

## 4.5 UN indices: PHDI and multidimensional poverty measures

The UN has developed several non-compensatory indices. The Human Development Index (HDI) uses the geometric mean. The planetary-adjusted HDI (PHDI) multiplies HDI by (1−Planetary Pressure index), penalising environmental pressures multiplicatively (United Nations Development Programme, 2026). The Multidimensional Poverty Index (MPI) uses a dual cutoff approach (OPHI and UNDP, 2025). The EU's AROPE uses maximum aggregation (at the individual level): a household is at risk if it falls into any of three categories.

These operate at aggregation (hence, out of scope for conditions ii, iii, vi). PHDI provides limited substitutability (iv). MPI and AROPE are highly non-compensatory. They do not address non-linearity (vii) and upper limits (viii).

## 4.6 Partial and non-compensatory MCDA: outranking methods and veto logic (Ricardo)

Outranking methods (ELECTRE, PROMETHEE, NAIADE) compare alternatives pairwise using concordance and discordance relations (Brans and Vincke, 1985; Munda, 1995; Roy, 1968). Rather than aggregating into a single score, they assess whether one alternative is at least as good as another. A key feature is veto thresholds: unacceptable performance on any criterion can prevent outranking regardless of other strengths. Social Multi-Criteria Evaluation (SMCE) integrates stakeholder perspectives (Munda, 2008, 2004). Multi-Criteria Mapping (MCM) keeps stakeholder perspectives separate (Stirling, 2006).

These operate at aggregation (hence, out of scope for conditions ii, iii, vi). They accommodate formative logic (i). They directly address limited substitutability (iv) and upper/lower limits (viii, ix) via veto thresholds. Limitations include reliance on preference parameters and partial rather than full non-compensability. Participatory variants enhance legitimacy but are resource intensive.

## 4.7 Data Envelopment Analysis (DEA)

DEA evaluates how well decision-making units transform inputs into outputs (Charnes et al., 1978; Farrell, 1957). It has been adapted to composite indicators under the label "Benefit-of-the-Doubt" (BoD) (Hasannasab et al., 2026) and applied to human development (Despotis, 2005), OECD Better Life indicators (Mizobuchi, 2014), and climate vulnerability (Walheer, 2024).

Rather than imposing fixed weights, DEA determined the weights endogenously from the data. For each country, the method identifies the most favourable set of weights for each unit and constructs an empirical frontier of the best performing units. Countries on the frontier score 1; others score below 1 depending on their distance from the frontier. Each country is judged under its most favourable weights — the "benefit of the doubt."

This flexibility is attractive for SIW, as different countries may pursue different development profiles. However, unrestricted flexibility is problematic. A country with a strong economic performance but poor environmental outcomes could receive zero weight for environmental indicators and be placed on the frontier, thus violating the premise that some dimensions are necessary and cannot be ignored. The solution is weight restrictions: lower bounds, ordinal constraints, or limits on each indicator's contribution (Cherchye et al., 2007; Nardo et al., 2005) allow to maintain flexibility and the normative structure of SIW.

DEA operates at aggregation (hence, out of scope for conditions ii, iii, vi). With restrictions, it addresses limited substitutability (iv), imbalance penalisation (v), and upper/lower limits (viii, ix).

Limitations: DEA weights are optimisation weights, not social preferences; results are sample-dependent; temporal comparison is challenging (Walheer, 2024).

## 4.8 Median-percentile aggregation (EU Resilience Dashboard and SIW Dashboard)

The European Commission's Resilience Dashboard normalises each indicator by its percentile rank in the pooled multiyear distribution (Benczur et al., 2023). Indicators are aggregated using the median of these percentiles — a highly non-compensatory summary. The 2005 pilot SIW Dashboard also adopted this methodology, though this may be revised in the future (Benczur et al., 2025).

This operates at normalisation and aggregation (hence, out of scope for conditions ii, iii, vi). It is highly non-compensatory (iv) and penalises imbalances (v) implicitly. It is robust to outliers. Limitations: percentile ranks depend on the reference sample; it does not address non-linearity (vii) and upper/lower limits (viii, ix).

# 5 LEARNING FROM OTHER FIELDS: CROSS-FIELD INSIGHTS FOR SIW AGGREGATION

The methods in Sections 3 and 4 come primarily from statistics, economics, and decision science. Other fields have long grappled with similar problems: combining multiple interacting factors when outcomes are non-linear, constraints are binding, and compensation is limited. This section draws insights from ecology, agronomy, chemistry, health, psychology, neuroscience, and machine learning. Each offers principles that inform SIW aggregation.

## 5.1 Limiting factors, saturation, and non-linear aggregation (ecology, agronomy, chemistry, health)

These fields model outcomes as the joint result of multiple constraints, with non-linearity and limited compensability. Three core principles emerge. First, the minimum or bottleneck principle. Liebig's law of the minimum holds that system performance is constrained by the scarcest essential input, not the average (Sands et al., 2009). Crop growth is limited by whichever nutrient is in shortest supply. Chemistry offers a parallel: the slowest step determines multi-step reaction dynamics. For SIW, a society with high income and education but severe environmental degradation may experience constrained wellbeing — analogous to a system limited by a single missing nutrient.

Second, co-limitation and multiplicative aggregation. When multiple factors jointly constrain outcomes, improvements in one dimension have little effect unless other constraints are relieved. Co-limitation frameworks use multiplicative aggregation (Seghouani et al., 2024). Low values strongly depress performance. This "AND-like" logic relates to geometric-mean aggregation.

Third, saturation and dose–response relationships. Agronomy uses dose–response functions where additional inputs raise performance at a decreasing rate (Mitscherlich-type curves; Dhanoa et al., 2022). Epidemiology formalises similar patterns with Hill equations or logistic functions (Goutelle et al., 2008). Composite endpoints in clinical research show a "minimum condition" logic: a single severe outcome often dominates the composite.

A concrete example is the Sustainable Wellbeing Index (SWI) (Costanza et al., 2016), which combines three pillars (Economic, Natural, Social) using a multiplicative, saturating function *SWI = f(E,N,S)*, which could be based on Michaelis-Menten kinetics (Cornish-Bowden, 2015):

$$SWI = L_{max} \cdot \frac{E}{k_e+E} \cdot \frac{N}{k_n+N} \cdot \frac{S}{k_s+S}, \qquad \text{(Eq. 5)}$$

where: $L_{max}$ is the maximum achievable SWI when all factors are simultaneously at their maximum; $k_e$, $k_n$ and $k_s$ are the "half saturation constant" for of E, N and S respectively – the value of E, N or S where the result of this term achieves 1/2 its maximum value. Each term saturates as the pillar increases; any pillar can become the limiting factor.

These principles operate at normalisation (non-linear transformations) and aggregation (minimum operator, multiplicative aggregation). They inform non-linear transformations (vii), limited substitutability (iv), and lower limits (ix). Limitations: health and agronomy models focus on individual-level or plot-level outcomes and may not capture systemic or planetary dimensions (e.g., cross-border spillovers, condition iii).

## 5.2 Positive psychology

Positive psychology conceptualises wellbeing as an emergent state from interacting psychological, social, and contextual factors (Ryan and Deci, 2001; Seligman and Csikszentmihalyi, 2000). Seligman's PERMA framework identifies five dimensions: positive emotion, engagement, relationships, meaning, and accomplishment (Seligman, 2011).

Research shows deficits in one domain cannot be fully compensated by surpluses elsewhere (Diener and Seligman, 2002; Ryff and Keyes, 1995) — a limiting-factor dynamic. The income–life satisfaction relationship is saturating (Easterlin, 1974; Jebb et al., 2018; Kahneman and Deaton, 2010). Falling below minimum social belonging produces disproportionately large declines (Baumeister and Leary, 1995; Keyes, 2002). Hedonic adaptation introduces dynamic, non-linear responses (Diener et al., 2006; Kubiszewski et al., 2020). Complementarities among dimensions further reinforce limitations of additive aggregation (Fredrickson, 2001; Ryff, 2014).

Positive psychology provides conceptual justification for non-linear and non-compensatory aggregation. It supports limited substitutability (iv), imbalance penalisation (v), non-linear transformations (vii), and lower limits (ix). Limitations: it focuses on individual-level wellbeing and does not directly address environmental sustainability (viii), spillovers (iii), or intertemporal aggregation (vi).

## 5.3 Neuroscience

Neuroscience reveals how the brain integrates multidimensional experiences into subjective wellbeing, offering three principles. First, a common neural currency. The brain maps money, health, and social rewards onto a single Subjective Value (SV) signal in the ventromedial prefrontal cortex and orbitofrontal cortex (Bartra et al., 2013; Levy and Glimcher, 2012).

Second, non-linear and interactive integration. The value of one attribute is modulated by another — physical pain dampens the utility of monetary gain (Park et al., 2011). This provides a neurobiological basis for non-substitutability.

Third, dynamic adaptation and reference dependence. Wellbeing is modelled as a dynamic aggregation of recent expectations and Reward Prediction Errors (RPE) with an exponential forgetting factor (Rutledge et al., 2014). The brain uses structure learning to reset reference points during radical shifts (Hunter and Gershman, 2018).

Neuroscience supports limited substitutability (iv), non-linear transformations (vii), and dynamic adaptation (relevant for condition vi). Limitations: findings are based on individual brain mechanisms, not societal aggregates. Implementation would require real-time happiness reports and longitudinal data, not routinely available. Neuroscience models tend to favour hedonic over eudaimonic components (Berridge and Kringelbach, 2015, 2011).

## 5.4 Machine Learning

Machine learning (ML) offers data-driven methods to learn non-linear relationships without pre-specifying functional forms. Neural Additive Models (NAMs) and Interpretable Generalized Additive Neural Networks (IGANN) learn flexible shape functions — capturing, for example, income satiation or the U-shaped age–wellbeing relationship (Agarwal et al., 2020; Kraus et al., 2024). Deep Lattice Networks (DLN) impose monotonicity constraints (You et al., 2017). SHAP values decompose each indicator's contribution to a prediction (Lundberg and Lee, 2017).

ML can identify saturation and non-linear stagnation, supporting non-linear transformations (vii). It can diagnose situations where overall wellbeing fails to rise because one essential dimension is insufficient — a form of limited substitutability (iv) diagnosis. SHAP values can contribute to weight derivation.

Limitations: most ML models are additive and do not naturally capture non-substitutability without manual design (Kraus et al., 2024). They focus on individual-level variance rather than collective structural dynamics (Pelt et al., 2024). Implementation requires high-quality longitudinal data. Finally, ML models reflect correlations, not normative importance.

Table 2 presents a summary of the methods analysed in terms of the level that they apply to and the key conditions for SIW they address.

**Table 2. Methods and conditions addressed**

| Method | Level | (i) | (ii) | (iii) | (iv) | (v) | (vi) | (vii) | (viii) | (ix) | (x) |
|---|---|---|---|---|---|---|---|---|---|---|---|
| Arithmetic mean (3.1), Geometric mean (3.2), LAM (3.3), PCA (3.4) | Aggregation | | | | | | | | | | |
| Consumption/income-based (4.1) | Measurement | | | ✓ | | | | | | | |
| Outcome-based regression (4.2) | Weight derivation | | | | | | | Partial | | | |
| Non-linear normalisation (4.3) | Normalisation | | | | | | | ✓ | | | |
| AMPI (4.4) | Normalisation + Aggregation | ✓ | ✓ | | | ✓ | ✓ | | | ✓ | |
| MSI (4.4) | Aggregation | ✓ | ✓ | | | ✓ | ✓ | | | ✓ | |
| PHDI/MPI/AROPE (4.5) | Aggregation | | | | ✓ | | | | ✓ | | |
| Outranking MCDA (ELECTRE, PROMETHEE, MCM) (4.6) | Aggregation | ✓ | | | ✓ | ✓ | ✓ | | ✓ | ✓ | |
| DEA (4.7) | Aggregation | ✓ | ✓ | | ✓ | ✓ | ✓ | | ✓ | ✓ | |
| Median-percentile (4.8) | Normalisation + Aggregation | | ✓ | | | ✓ | | | | | |
| Ecology/agronomy/chemistry/health fields (minimum operator, multiplicative, co-limitation proxy, saturation variant) (5.1) | Normalisation + Aggregation | ✓ | ✓ | | ✓ | ✓ | ✓ | ✓ | | ✓ | |
| Positive psychology (5.2) | Conceptual | ✓ | ✓ | | ✓ | ✓ | ✓ | ✓ | | ✓ | |
| Neuroscience (5.3) | Conceptual | ✓ | | | ✓ | | | ✓ | | | |
| Machine Learning (5.4) | Weight derivation + Diagnostics | | | | | | | ✓ | | | |

Acronyms: LAM = Linear Additive Model; PCA = Principal Component Analysis; AMPI = Adjusted Mazziotta-Pareto Index; MSI = Multidimensional Synthesis Indicator; PHDI = Planetary Pressures adjusted Human Development Index; MPI = Multidimensional Poverty Index; MCDA = multicriteria decision analysis; MCM – Multicriteria Mapping; DEA = Data Envelopment Analysis.

# 6 ILLUSTRATIVE APPLICATION: COMPARING AGGREGATION METHODS

## 6.1 Aim and methods

This illustrative application compares aggregation methods for SIW and shows how alternative assumptions about compensability, bottlenecks, and non-linearity affect country rankings.

### 6.1.1 Illustrative dataset and normalisation

We constructed a fictitious dataset for 10 countries and five SIW domains: $CO_2$ emissions per capita, material footprint per capita, life expectancy, education, and income (GDP per capita, PPP). The dataset was designed to highlight trade-offs and extreme behaviours: country C has environmental dimensions near zero; country I has social dimensions near zero. All domains were normalised to a 0–1 scale with higher values indicating better performance. Table 3 reports the mean normalised values; minimum, median, and maximum values are provided in Supplementary Table S1 for methods requiring distributional information (e.g., MCM, median-percentile method).

**Table 3. Mean normalised scores for SIW domains**

| Country | $CO_2$ emissions | Material flows | Life expectancy | Education | Income |
|---|---|---|---|---|---|
| A | 0.68 | 0.60 | 1.00 | 1.00 | 1.00 |
| B | 0.87 | 0.84 | 0.89 | 0.90 | 0.65 |
| C | 0.00 | 0.00 | 0.74 | 0.70 | 0.93 |
| D | 0.53 | 0.48 | 0.84 | 0.80 | 0.54 |
| E | 0.97 | 0.94 | 0.95 | 0.98 | 0.43 |
| F | 0.93 | 0.80 | 0.95 | 0.93 | 0.34 |
| G | 0.77 | 0.68 | 0.63 | 0.63 | 0.21 |
| H | 0.72 | 0.64 | 0.47 | 0.45 | 0.27 |
| I | 1.00 | 1.00 | 0.00 | 0.00 | 0.00 |
| J | 0.66 | 0.52 | 0.42 | 0.40 | 0.19 |

### 6.1.2 Aggregation methods compared

We selected 13 methods spanning the families reviewed in Sections 3-5: arithmetic average, geometric average, linear additive model, PHDI, minimum operator, co-limitation proxy, saturation variant, outranking MCDAs: PROMETHEE and MCM, DEA, PCA, the median-percentile methods, and a neuroscience-machine learning index.

*Compensatory methods (baseline)*

Arithmetic average and geometric average both use Eq. 1, with p=1 and 0, respectively. $w_i$ were set to 1. For LAM, we have used Eq. 2 for $w_i$ defined in Table 4.

**Table 4. Weights used for LAM, PROMETHEE and MCM**

| Component | LAM and PROMETHEE | MCM |
|---|---|---|
| $CO_2$ emissions | 0.30 | 0.2-0.4 |
| Material footprint | 0.20 | 0.1-0.3 |
| Life expectancy | 0.20 | 0.1-0.3 |

| Component | LAM and PROMETHEE | MCM |
|---|---|---|
| Education | 0.15 | 0.1-0.2 |
| Income | 0.15 | 0.1-0.2 |

For the PHDI aggregation, we took the arithmetic mean of the two environmental domains, the arithmetic mean of the three social domains, then combined the two blocks using a geometric mean.

*Partial and non-compensatory methods*

For the Minimum operator (Liebig-type bottleneck), takes the weakest component $SIW = min_i(x_i)$. The co-limitation proxy combines bottleneck and average: $(min\ (x_i)+mean(x_i))/2$. The saturation variant applies a square root transformation to each component before averaging.

*Outranking MCDA methods*

PROMETHEE rankings were computed in Visual PROMETHEE using the weights in Table 4 and default preference-function settings. Multi-Criteria Mapping (MCM) differs from LAM in two ways: weights are represented as ranges (Table 4) to capture stakeholder heterogeneity, and veto criteria can exclude options that fall below a threshold on any criterion. We evaluated two weight configurations (MCM1, MCM2) and implemented an environmental veto rule (exclude if either environmental component < 0.2).

*Flexible weighting (DEA)*

We constructed a Benefit-of-the-Doubt (BoD) index using output-oriented DEA under variable returns to scale, choosing non-negative weights *wj* to maximise Eq. 6 subject to the benchmarking constraints (Eq. 7) for all countries (*r*).

$$\boldsymbol{max_{w\geq 0}}\ \sum_{j=1}^{5} \boldsymbol{w_j\, y_{ij}} \qquad \text{(Eq. 6)}$$

$$\sum_{j=1}^{5} \boldsymbol{w_j} \leq \boldsymbol{1} \qquad \text{(Eq. 7)}$$

To enforce SIW-consistent weighting, we imposed an environmental share constraint (Eq. 8) and lower bounds (Eq. 9).

$$\boldsymbol{w_{CO2}} + \boldsymbol{w_{MF}} \geq \boldsymbol{0.5} \qquad \text{(Eq. 8)}$$

$$\boldsymbol{w_j} \geq \boldsymbol{0.05}\ , \qquad \text{(Eq. 9)}$$

BoD scores lie in [0,1]; a score of 1 indicates the country attains the empirical best-practice frontier under at least one admissible weighting scheme. We implemented this model using the GRG non-liner method in the *solver* package in Microsoft Excel.

*Data-driven methods (diagnostic)*

PCA (first principal component) is included as a diagnostic comparator — a demonstration of what not to do for SIW, given its reflective assumptions and potential for negative loadings. The median-percentile method (JRC/EC Resilience Dashboard and SIW Dashboard) normalises by percentile rank and aggregates by the median.

*Illustrative hybrid method (Neuro-ML)*

For the Neuro-ML index, we imposed non-linear transformations (power function for environmental indicators, logistic for life expectancy and education, logarithmic for income), interaction terms, and reward-prediction-error adjustments. These are illustrative assumptions; in practice, such forms would be estimated from data.

#### 6.1.3 Comparing method outputs

To quantify agreement between methods, we computed Spearman rank correlations ρ across the country orderings. Spearman's ρ depends only on ranks, making it invariant to differences in score scale (e.g., PROMETHEE net flows vs. additive indices). Ties were assigned average ranks.

## 6.2 Results

#### 6.2.1 Method families and concordance: what the clusters reveal about SIW conditions

Spearman rank correlations (Figure 1) reveal three distinct clusters. Interpreted through the lens of the nine SIW conditions (Table 1), each cluster reflects a different approach to key principles.

First, compensatory and averaging-based methods (arithmetic average, LAM, geometric mean, PHDI) are highly concordant, with correlations up to ρ = 0.98. These methods reward high average achievement, even when performance is uneven across dimensions, but violate several core SIW conditions. Specifically, they assume full substitutability (condition i), impose no penalty for imbalances (condition ii), are linear (condition iii), and treat upper and lower limits (conditions iv, v) as optimisable rather than as hard constraints. DEA, with its imposed weight restrictions, partially addresses condition (i) and shows strong alignment with this cluster (ρ = 0.83–0.88 with averaging methods), confirming that constrained flexible weighting largely preserves the ordering of compensatory approaches while embedding normative constraints.

Second, limiting-factor methods (minimum operator, co-limitation proxy, saturation variant) are highly concordant among themselves (ρ > 0.83) and show moderate-to-high agreement with the compensatory cluster (ρ = 0.61–0.92). This results in similar broad ordering but systematic re-ranking of countries with unbalanced profiles. Limiting-factor methods better address limited substitutability (i) and penalisation of imbalances (ii). The saturation variant explicitly addresses non-linear transformations (iii) through concave transformations, and all three methods implicitly respect lower limits (v) by penalising extreme shortfalls. DEA correlates strongly with these methods as well (ρ = 0.77–0.88), reflecting its sensitivity to imbalances through weight restrictions.

Third, outranking-based MCDA methods (PROMETHEE, MCM) are highly concordant among themselves (ρ = 0.94–0.96) but only moderately correlated with compensatory methods (ρ = 0.54–0.60). This lower concordance implies that changing the aggregation paradigm—from averaging to pairwise outranking with veto logic—produces meaningfully different cross-country comparisons. Outranking methods directly address limited substitutability (i) through veto thresholds and can enforce upper limits (iv)—as MCM does in this case—and lower limits (v) as hard constraints. They also accommodate formative measurement logic (vi) and impose no correlation requirement (vii). Notably, DEA correlates more strongly with outranking methods (ρ = 0.87–0.90) than with the compensatory cluster, suggesting that flexible weighting and frontier-based benchmarking share some conceptual ground with pairwise comparison logic.

PCA stands apart. Its negative correlations with most methods (ρ as low as -0.67) reflect its fundamental misalignment with SIW's formative measurement model (condition vi). This is because PCA focuses on what varies most, rather than what matters most. We return to this diagnostic case in Section 6.2.4.

| | Arith. avg | LAM | Geom | PHDI | Min | Co-lim | Sat | PMT | MCM1 | MCM2 | DEA | Neuro-ML | Median | PCA |
|---|---|---|---|---|---|---|---|---|---|---|---|---|---|---|
| **Arith. avg** | 1.00 | 0.94 | 0.90 | 0.82 | 0.75 | 0.79 | 0.86 | 0.60 | 0.54 | 0.59 | 0.83 | 0.22 | 0.44 | -0.14 |
| **LAM** | 0.94 | 1.00 | 0.81 | 0.83 | 0.76 | 0.84 | 0.83 | 0.84 | 0.84 | 0.83 | 0.83 | 0.18 | 0.43 | -0.29 |
| **Geom** | 0.90 | 0.81 | 1.00 | 0.98 | 0.74 | 0.82 | 0.64 | 0.79 | 0.81 | 0.78 | 0.88 | 0.33 | 0.59 | -0.65 |
| **PHDI** | 0.82 | 0.83 | 0.98 | 1.00 | 0.67 | 0.78 | 0.66 | 0.82 | 0.77 | 0.77 | 0.88 | 0.25 | 0.55 | -0.67 |
| **Min** | 0.75 | 0.76 | 0.74 | 0.67 | 1.00 | 0.92 | 0.65 | 0.54 | 0.47 | 0.49 | 0.77 | 0.37 | 0.56 | -0.13 |
| **Co-lim** | 0.79 | 0.84 | 0.82 | 0.78 | 0.92 | 1.00 | 0.83 | 0.64 | 0.58 | 0.60 | 0.81 | 0.48 | 0.72 | 0.15 |
| **Sat** | 0.86 | 0.83 | 0.64 | 0.66 | 0.65 | 0.83 | 1.00 | 0.64 | 0.71 | 0.69 | 0.88 | 0.50 | 0.55 | -0.02 |
| **PMT** | 0.60 | 0.84 | 0.79 | 0.82 | 0.54 | 0.64 | 0.64 | 1.00 | 0.96 | 0.94 | 0.90 | 0.26 | 0.50 | 0.19 |
| **MCM1** | 0.54 | 0.84 | 0.81 | 0.77 | 0.47 | 0.58 | 0.71 | 0.96 | 1.00 | 0.95 | 0.87 | 0.24 | 0.48 | 0.18 |
| **MCM2** | 0.59 | 0.83 | 0.78 | 0.77 | 0.49 | 0.60 | 0.69 | 0.94 | 0.95 | 1.00 | 0.87 | 0.24 | 0.49 | 0.15 |
| **DEA** | 0.83 | 0.83 | 0.879 | 0.879 | 0.77 | 0.806 | 0.879 | 0.903 | 0.867 | 0.867 | 1.00 | 0.806 | 0.842 | -0.006 |
| **Neuro-ML** | 0.22 | 0.18 | 0.33 | 0.25 | 0.37 | 0.48 | 0.50 | 0.26 | 0.24 | 0.24 | 0.81 | 1.00 | 0.13 | -0.49 |
| **Median** | 0.44 | 0.43 | 0.59 | 0.55 | 0.56 | 0.72 | 0.55 | 0.50 | 0.48 | 0.49 | 0.84 | 0.13 | 1.00 | 0.43 |
| **PCA** | -0.14 | -0.29 | -0.65 | -0.67 | -0.13 | 0.15 | -0.02 | 0.19 | 0.18 | 0.15 | -0.01 | -0.49 | 0.43 | 1.00 |

**Figure 1. Spearman rank correlations between aggregation methods**

Abbreviations: Arith. avg = arithmetic average; LAM = weighted linear additive model; Geom = geometric mean; PHDI = PHDI-type aggregation; Min = minimum operator; Co-lim = co-limitation proxy; Sat = saturation variant; PMT = PROMETHEE, an outranking MCDA; MCM1/2 = multi-criteria mapping variants, a non/partially compensatory MCDA; DEA = Data Envelopment Analysis; Neuro-ML = neuroscience/machine-learning index; PCA = principal component analysis.

### 6.2.2 Rank sensitivity: how condition adherence shapes rankings

Aggregation choices produce marked differences in country rankings (Table 5). The largest dispersion occurs for Country C, which has near-zero environmental performance but high income. It ranks first under PCA but last or near-last under most limiting-factor and outranking methods. This extreme divergence illustrates how different methods handle core SIW conditions:

- PCA (rank 1) violates formative logic (vi) and respect for upper limits (iv), treating high income as sufficient to offset environmental failure.
- MCM excludes C entirely, demonstrating how veto thresholds operationalise respect for upper limits (iv) as a hard constraint.
- All remaining methods (apart from the neuroscience-machine learning index and the median percentile) rank C moderately high (ranks 8–10).

Countries with balanced profiles (G, H, J) show narrow rank ranges (2 positions or less). Disagreement concentrates on boundary cases with extreme shortfalls in one dimension—precisely where assumptions about compensation and limits matter most.

**Table 5. Country rankings across aggregation methods (1 = best; 10 = worst)**

| Country | Arith. avg | LAM | Geom | PHDI | Min | Co-lim | Sat | PMT | MCM | DEA | Neuro-ML | Median | PCA | Range of rankings [a] |
|---|---|---|---|---|---|---|---|---|---|---|---|---|---|---|
| A | 1 | 3 | 1 | 3 | 2 | 2 | 1 | 4 | 1 - 3 | 5 | 3 | 1 | 2 | 4 |
| B | 3 | 2 | 2 | 2 | 1 | 1 | 3 | 3 | 3 - 2 | 3 | 2 | 4 | 4 | 3 |
| C | 8 | 10 | 9 | 9 | 9 | 9 | 9 | 9 | Excl. | 10 | 7 | 6 | 1 | 9[b] |
| D | 5 | 5 | 5 | 5 | 3 | 5 | 5 | 7 | 6 | 9 | 5 | 5 | 3 | 6 |
| E | 2 | 1 | 3 | 1 | 4 | 3 | 2 | 1 | 2 - 1 | 1 | 1 | 2 | 6 | 5 |
| F | 4 | 4 | 4 | 4 | 5 | 4 | 4 | 2 | 4 | 2 | 4 | 3 | 5 | 3 |
| G | 6 | 6 | 6 | 6 | 7 | 6 | 6 | 6 | 5 | 6 | 6 | 7 | 7 | 2 |
| H | 7 | 7 | 7 | 7 | 6 | 7 | 7 | 7 | 7 | 7 | 8 | 7 | 8 | 2 |
| I | 10 | 8 | 9 | 9 | 9 | 10 | 10 | 5 | 9 - 8 | 4 | 10 | 10 | 10 | 6 |
| J | 9 | 9 | 8 | 8 | 8 | 8 | 8 | 10 | 8 - 9 | 8 | 9 | 9 | 9 | 2 |

Abbreviations: Arith. avg = arithmetic average; LAM = weighted linear additive model; Geom = geometric mean; PHDI = PHDI-type aggregation; Min = minimum operator; Co-lim = co-limitation proxy; Sat = saturation variant; PMT = PROMETHEE, an outranking MCDA; MCM1/2 = multi-criteria mapping variants, a non/partially compensatory MCDA; DEA = Data Envelopment Analysis; Neuro-ML = neuroscience/machine-learning index; PCA = principal component analysis.
(a) Rank range is computed as the difference between the best (minimum) and worst (maximum) rank observed across methods. For MCM, the interval [best–worst] is shown.
(b) Does not include "Excluded".

### 6.2.3 Penalising shortfalls: what non-compensatory methods reveal about conditions (i), (ii), (iv), and (v)

Fully compensatory approaches (arithmetic average, LAM) allow high income to offset near-zero environmental performance. Country C benefits from this logic. Limiting-factor methods (minimum, co-limitation) and outranking methods with veto rules reverse this ranking, placing Country C near the bottom.

This difference reflects how different methods treat four interconnected conditions:

- Limited substitutability (i): Compensatory methods allow full substitution; limiting-factor methods restrict it.
- Penalisation of imbalances (ii): Compensatory methods ignore imbalances; limiting-factor methods penalise them.
- Respect for upper limits (iv): Compensatory methods treat planetary boundaries as trade-offs; MCM's veto rule enforces them as hard constraints.
- Respect for lower limits (v): The minimum operator enforces a de facto lower limit by letting the weakest dimension dominate.

For SIW, where planetary boundaries are necessary conditions and basic needs are non-substitutable, the partial or non-compensatory logic is more defensible. The ranking reversal for Country C reflects competing normative assumptions about whether environmental degradation can be compensated by economic prosperity.

### 6.2.4 Boundary cases and diagnostic warnings: when methods violate conditions (vi), (i), and (vii)

PCA assigns negative loadings to environmental variables due to their negative correlation with socio-economic indicators, capturing a "socio-economic development–environment trade-off axis" rather than an SIW-aligned composite. This illustrates a known limitation: for formative constructs where domains are not expected to correlate, PCA answers the wrong question. In principle, PCA could be applied to a restricted subset of indicators expected to share a common underlying domain—where positive loadings would be anticipated. For example, one could use PCA for the socio-economic and environmental domains separately and then aggregate the two dimensions with a different method. However, when indicators from different wellbeing domains (e.g., environmental and socio-economic) are mixed, negative correlations are common, and a single latent variable is insufficient to capture the construct. This reinforces that PCA is not suitable as a standalone SIW aggregator, particularly for formative constructs where domains are not expected to correlate. We include PCA here not as a recommended method, but as a diagnostic warning.

Country I (strong environment, near-zero social outcomes) ranks relatively high under PROMETHEE (rank 5) compared to most other methods (ranks 8–10). This occurs because, without explicit veto thresholds, poor performance on social criteria need not be decisive if a country strongly outranks others on heavily weighted criteria (here, the environmental block). This emphasises that limited substitutability (i) and respect for lower limits (v) are not automatically enforced by outranking logic—they depend on preference-function settings and the presence or absence of vetoes. These choices operationalise the analyst's view of compensability.

#### 6.2.5 Limitations of the illustrative exercise

This exercise is stylised. It uses a small sample (10 countries) and a single cross-section, limiting temporal robustness. The dataset was constructed to include strong trade-offs; real-world data may differ. Several method implementations require modelling choices (weight vectors, preference functions, veto rules) that could change rankings. Most importantly, intertemporal aggregation (condition x) is not addressed by any method reviewed. None of the methods considered here account for capital depletion, future wellbeing, or discounting—a gap that reflects the conceptual difficulty of this condition rather than a limitation of the illustrative exercise itself.

# 7 DISCUSSION

## 7.1 Summary of findings

This paper derived nine conditions for SIW from needs theory and strong sustainability (Section 2) and reviewed aggregation methods against these. Our analysis yielded three main findings. First, conventional aggregation methods are fundamentally ill-suited to SIW. Weighted and unweighted arithmetic means—the default in many wellbeing indicators—assume full substitutability, linearity, and the absence of environmental or social limits. These assumptions conflict directly with the core nature of SIW, where deficits in health or ecological integrity cannot be fully offset by income gains, and where planetary boundaries act as hard constraints rather than optimisable variables.

Second, non-compensatory and partially compensatory methods better reflect SIW's structure, yet no single method satisfies all nine conditions. Outranking methods (PROMETHEE, MCM) limit substitutability through veto thresholds. Penalty-based indices (AMPI, MSI) penalise imbalances and allow non-linear transformations. DEA with weight restrictions offers flexible benchmarking but requires careful design. Neuroscience and machine learning provide valuable non-linear insights but are data-intensive and individual-focused. Each method addresses a different subset of the nine conditions; none covers them all.

Third, aggregation choices materially change country rankings and, with them, policy narratives. In our illustrative example (Section 6), Country C—with near-zero environmental performance but high income—ranks first under PCA but last or near last under most non-compensatory methods. This is not a technical artefact. It reflects competing normative assumptions about weak and strong sustainability: whether environmental degradation can be compensated by economic prosperity (Benczur et al., 2025; Costanza et al., 1997). For SIW, where planetary boundaries are necessary conditions (condition iv) and human needs are non-substitutable (condition i), non-compensatory logic is more defensible. Compensatory methods cluster tightly together ($\rho$ up to 0.98), while outranking methods are only moderately correlated with averaging ($\rho \approx 0.55$–$0.60$). Changing the aggregation paradigm changes the policy story.

## 7.2 Implications

These findings carry several implications for beyond-GDP measurement and policy. For monitoring and reporting, international organisations and national statistical offices that currently rely on arithmetic means or PCA should reconsider those choices. The UN's PHDI, the EU's median-percentile approach, and outranking MCDA already embed partial or non-compensatory logic, demonstrating that moving beyond linear averaging is both theoretically justified and practically feasible. Future SIW dashboards should adopt aggregation methods that penalise imbalances, respect environmental ceilings, and limit substitutability across dimensions.

For policy design, the shift to partial or non-compensatory aggregation changes incentives. Under compensatory methods, a country can trade off poor environmental performance against high income—a logic that rewards lopsided development. Under limiting-factor methods, critical shortfalls cannot be ignored. This redirects policy attention toward bottlenecks and environmental ceilings, aligning with the strong sustainability principle that natural capital is non-substitutable. Penalty-based indices that penalise imbalances encourage balanced development across wellbeing dimensions rather than maximising a few at the expense of others.

For participatory and deliberative processes, methods such as MCM and SMCE allow stakeholders to express veto thresholds and weight ranges, making normative trade-offs explicit. This transparency is valuable for legitimate policy decisions. However, these processes are resource-intensive and difficult to replicate consistently across countries and over time, limiting their applicability for routine cross-country monitoring.

Taken together, non-compensatory elements can help shift policy incentives toward addressing bottlenecks and respecting ecological ceilings, while inequality-sensitive transformations prioritise inclusiveness. The tools exist; the question is whether they will be used.

## 7.3 Novel contributions

This paper makes three novel contributions to the literature on wellbeing measurement. First, we synthesise and apply a coherent set of nine conditions for SIW aggregation, drawn from needs theory and strong sustainability (Section 2, Table 1). These conditions provide a normative benchmark against which existing and future aggregation methods can be assessed.

Second, we provide an illustrative comparison of 13 aggregation methods on a common dataset (Section 6), showing how different assumptions about compensability, non-linearity, and thresholds produce different country rankings. To our knowledge, this is the first systematic illustration of such a wide range of methods for SIW.

Third, we bridge across fields—ecology, agronomy, chemistry, health, positive psychology, neuroscience, and machine learning (Section 5)—to identify aggregation principles (minimum, co-limitation, saturation) that are rarely applied in wellbeing measurement but offer valuable insights for SIW.

The UN High-Level Expert Group on Beyond GDP (2026) recently called for urgent action on better wellbeing measures. Notably, the HLEG did not propose a single composite index structure based on its dashboard, but it recommended establishing a scientific committee to further develop headline aggregate indicators that transparently and rigorously capture the key dimensions of progress. Our paper directly supports that initiative by reviewing and illustrating aggregation methods that could underpin such headline indicators.

## 7.4 Limitations and future work

This study has a few limitations, which also point to directions for future research.

First, our illustrative dataset is small (10 countries), fictitious, and cross-sectional. Real-world SIW monitoring would require larger samples, multiple time periods, and real data. Future work should replicate this comparison on empirical datasets, including sensitivity analyses over key parameters such as weights, thresholds, and veto rules.

Second, none of the methods reviewed address intertemporal aggregation (condition x). How should we weigh current versus future wellbeing? How should we account for depletion of natural or social capital? What discount rate—if any—should apply to future SIW? As Costanza et al. (2021) argue, different types of capital (built, human, social, natural) have inherently different characteristics and should be discounted (or their dividends) at different rates, if at all. A pluralistic discounting approach could in principle be integrated into a multi-period SIW composite indicator, but no existing method we reviewed does so.

Third, we have not addressed uncertainty propagation, sensitivity to data gaps, or robustness to alternative normalisation procedures. Future SIW indices should incorporate uncertainty and sensitivity analysis as standard practice (Saisana et al., 2005).

Fourth, this paper reviews methods but does not propose a final SIW production function. Such a function will require explicit saturation and threshold parameters (derived from needs literature or empirical dose–response), a non-compensatory aggregation rule, distribution-sensitive transformations, and a way to integrate subjective wellbeing alongside objective conditions. This remains the next frontier.

These gaps point to several research priorities. On method development, future work should design hybrid aggregation methods, starting from a formative logic, and that combine saturation and threshold transformations at the normalisation stage with partial or non-compensatory aggregation rules. On empirical parameters, saturation thresholds—such as income satiation points or minimum social connection levels—and half-saturation constants should be estimated from longitudinal, multi-country datasets rather than imposed arbitrarily. On participatory methods, research should explore whether deliberative processes at the national level can produce stable, reusable weight ranges and veto rules across cultural contexts. On intertemporal aggregation—the most neglected condition—future work should explore whether composite indicators can incorporate capital depletion as a penalty, or whether entirely new frameworks are needed to account for future wellbeing.

# 8 RECOMMENDATIONS

Based on our findings, we offer the following recommendations for policymakers, statistical offices, and researchers.

For immediate application, international and national bodies constructing wellbeing dashboards should consider four aspects. First, use median-based or minimum-based aggregation as highly non-compensatory and robust alternatives, particularly for diagnostic monitoring of critical shortfalls (e.g., the EU Resilience Dashboard approach). Second, avoid using simple arithmetic means or PCA as the sole aggregation method. These assume full compensation and linearity, inconsistent with SIW. If used, they should be restricted to lower levels of the conceptual framework (domains or sub-domains) and accompanied by non-compensatory complements. Third, adopt penalty-based indices such as AMPI or MSI, which penalise imbalances and are already used in official statistics (e.g., Italy's well-being index). These methods are transparent, reproducible, and compatible with formative measurement logic. Fourth, incorporate veto thresholds for dimensions that are truly necessary conditions, such as planetary boundaries (condition iv) or minimum health standards (condition v). Outranking methods like PROMETHEE or MCM can implement this logic, as can simpler rule-based approaches (e.g., excluding countries or assigning a minimum score below a certain environmental threshold).

The UN High-Level Expert Group on Beyond GDP has called for urgent action. Our review shows that better tools already exist. The challenge is no longer technical feasibility—it is political will and institutional capacity. We therefore recommend that the UN, the OECD, the European Commission, and national statistical offices support: (1) adopting a hybrid aggregation framework that combines saturation normalisation, imbalance penalisation, and veto logic for critical dimensions; (2) piloting this framework on real data across multiple countries and years; and (3) publishing transparent sensitivity analyses and uncertainty intervals alongside point estimates.

The cost of delay is borne today and by future generations. Many ingredients are already available. What is needed now is a practical assessment of which can be applied, and the courage to make an eventual proposal.

# 9 CONCLUSIONS

The pursuit of sustainable and inclusive wellbeing (SIW) has moved from academic debate to urgent policy action. The UN High-Level Expert Group on Beyond GDP (2026) called for measuring progress in terms of equitable, inclusive, and sustainable wellbeing, concluding that delay is no longer an option. Yet the HLEG did not propose how to aggregate its dashboard into a composite index—an omission that this paper aims to address .

Our review leads to four conclusions. First, conventional aggregation methods are actively misleading for SIW. Arithmetic means assume full substitutability, linearity, and no environmental or social limits. These assumptions conflict with SIW, where deficits in health or ecological integrity cannot be offset by income gains, and planetary boundaries are hard constraints.

Second, no single method satisfies all nine conditions that SIW requires. Penalty-based indices (AMPI, MSI) penalise imbalances. Outranking methods (PROMETHEE, MCM) limit substitutability via veto thresholds. DEA with weight restrictions offers flexible benchmarking. Neuroscience, machine learning, and insights from ecology and health provide valuable non-linear insights but are not standalone solutions. Each method addresses a subset of conditions; none covers them all.

Third, aggregation choices materially change country rankings and policy narratives. In our illustrative example, Country C—with near-zero environmental performance but high income—ranks first under PCA but last or near last under most non-compensatory methods. Compensatory methods cluster tightly together ($\rho$ up to 0.98), while outranking methods are only moderately correlated with averaging ($\rho \approx 0.55$–$0.60$). Changing the aggregation paradigm changes the policy story.

Fourth, a future SIW composite indicator will require combining methods across levels. Saturation and threshold effects are best addressed at normalisation. Limited substitutability and veto logic require non-compensatory aggregation. Penalising imbalances calls for methods like AMPI or MSI. Inclusiveness and cross-border spillovers require measurement-level choices. Intertemporal aggregation (which refers to broader aspects such as future wellbeing estimation and integration with present wellbeing) remains the most neglected condition.

The tools to move beyond linear averaging already exist. The UN's PHDI, the EU's median-percentile approach, and outranking MCDA demonstrate that non-compensatory logic is both justified and feasible. We therefore call on the UN, OECD, European Commission, and national statistical offices to support the adoption of a hybrid composite index (saturation normalisation, imbalance penalisation, veto logic), pilot it on real data, and invest in research on intertemporal aspects. The cost of delay is borne today and by future generations. We have the ingredients. What is needed now is the courage to use them.

## 10 DECLARATION OF COMPETING INTERESTS

No competing interests to report

## 11 FUNDING SOURCES

the views expressed are those of the authors and do not reflect those of their respective organisations.

This project has received funding from the European Union's Horizon Europe research and innovation programme under grant agreement No 101132524 (MERGE – Measuring what matters: Improving usability and accessibility of policy frameworks and indicators for multidimensional well-being through collaboration). The work of RSV has been supported by the European Union's Horizon Europe research and innovation programme (Maps project, grant agreement nr. 10113791). Views and opinions expressed are however those of the authors only and do not necessarily reflect those of the European Union or the European Research Executive Agency. Neither the European Union nor the European Research Executive Agency can be held responsible for them.

## 12 ACKNOWLEDGEMENTS

We would like to acknowledge JRC for offering the space for groups meeting and discussions on the paper. We would like to thank Panagiotis Ravanos (JRC) for very useful discussions on the topic.

## 13 DECLARATION OF GENERATIVE AI AND AI-ASSISTED TECHNOLOGIES IN THE WRITING PROCESS

During the preparation of this work the authors used ChatGPT and DeepSeek to improve the readability and language of the article. After using this tool/service, the authors reviewed and edited the content as needed and take full responsibility for the content of the published article.

## 14 REFERENCES

Agarwal, R., Melnick, L., Frosst, N., Zhang, X., Lengerich, B., Caruana, R., Hinton, G., 2020. Neural Additive Models: Interpretable Machine Learning with Neural Nets. https://doi.org/10.48550/ARXIV.2004.13912

Banville, C., Landry, M., Martel, J.-M., Boulaire, C., 1998. A stakeholder approach to MCDA. Syst. Res. Behav. Sci. 15, 15–32. https://doi.org/10.1002/(SICI)1099-1743(199801/02)15:1%3C15::AID-SRES179%3E3.0.CO;2-B

Bartra, O., McGuire, J.T., Kable, J.W., 2013. The valuation system: A coordinate-based meta-analysis of BOLD fMRI experiments examining neural correlates of subjective value. NeuroImage 76, 412–427. https://doi.org/10.1016/j.neuroimage.2013.02.063

Baumeister, R.F., Leary, M.R., 1995. The need to belong: Desire for interpersonal attachments as a fundamental human motivation. Psychol. Bull. 117, 497–529. https://doi.org/10.1037/0033-2909.117.3.497

Benczur, P., Boskovic, A., Giovannini, E., Pagano, A., Sandor, A.-M., 2025. Measuring sustainable and inclusive wellbeing: a multidimensional dashboard approach. Publications Office of the European Union, LU.

Benczur, P., Cariboni, J., Joossens, E., Le Blanc, J., Menyhert, B., Piriu, A., 2023. Monitoring resilience in the EU: evidence from the 2023 edition of the Resilience Dashboards. Publications Office of the European Union, Luxembourg.

Berridge, K.C., Kringelbach, M.L., 2015. Pleasure Systems in the Brain. Neuron 86, 646–664. https://doi.org/10.1016/j.neuron.2015.02.018

Berridge, K.C., Kringelbach, M.L., 2011. Building a neuroscience of pleasure and well-being. Psychol. Well- Theory Res. Pract. 1, 3. https://doi.org/10.1186/2211-1522-1-3

Biggeri, M., Mauro, V., 2018. Towards a more 'Sustainable' Human Development Index: Integrating the environment and freedom. Ecol. Indic. 91, 220–231. https://doi.org/10.1016/j.ecolind.2018.03.045

Boskovic, A., Sandor, A., Benczúr, P., Zec, S., 2026. Towards a unified framework for measuring sustainable and inclusive wellbeing in the EU. Humanit. Soc. Sci. Commun.

Brans, J.P., Vincke, P., 1985. A Preference Ranking Organisation Method: (The PROMETHEE Method for Multiple Criteria Decision-Making). Manag. Sci. 31, 647–656.

Brundtland, G.H., Khalid, M., Agnelli, S., Al-Athel, S., Chidzero, B., 1987. Our common future. N. Y. 8.

Burger, M.J., Courchesne, S., Greyling, T., O'Connor, K., Rossouw, S., Sarracino, F., Veenhoven, R., 2026. Burger, M.J., Courchesne, S., Greyling, T. et al. Using Subjective Well-being as a Headline Indicator in Dashboards to Track Human Progress. Applied Research Quality Life (2026). Appl. Res. Qual. Life. https://doi.org/https://doi.org/10.1007/s11482-026-10576-9

Charnes, A., Cooper, W.W., Rhodes, E., 1978. Measuring the efficiency of decision making units. Eur. J. Oper. Res. 2, 429–444. https://doi.org/10.1016/0377-2217(78)90138-8

Cherchye, L., Moesen, W., Rogge, N., Puyenbroeck, T.V., 2007. An Introduction to 'Benefit of the Doubt' Composite Indicators. Soc. Indic. Res. 82, 111–145. https://doi.org/10.1007/s11205-006-9029-7

Cornish-Bowden, A., 2015. One hundred years of Michaelis–Menten kinetics. Perspect. Sci. 4, 3–9. https://doi.org/10.1016/j.pisc.2014.12.002

Costanza, R., d'Arge, R., De Groot, R., Farber, S., Grasso, M., Hannon, B., Limburg, K., Naeem, S., O'Neill, R.V., Paruelo, J., Raskin, R.G., Sutton, P., Van Den Belt, M., 1997. The value of the world's ecosystem services and natural capital. Nature 387, 253–260. https://doi.org/10.1038/387253a0

Costanza, R., Daly, L., Fioramonti, L., Giovannini, E., Kubiszewski, I., Mortensen, L.F., Pickett, K.E., Ragnarsdottir, K.V., De Vogli, R., Wilkinson, R., 2016. Modelling and measuring sustainable wellbeing in connection with the UN Sustainable Development Goals. Ecol. Econ. 130, 350–355. https://doi.org/10.1016/j.ecolecon.2016.07.009

Costanza, R., Kubiszewski, I., Stoeckl, N., Kompas, T., 2021. Pluralistic discounting recognizing different capital contributions: An example estimating the net present value of global ecosystem services. Ecol. Econ. 183, 106961. https://doi.org/10.1016/j.ecolecon.2021.106961

Despotis, D.K., 2005. Measuring human development via data envelopment analysis: the case of Asia and the Pacific. Omega 33, 385–390. https://doi.org/10.1016/j.omega.2004.07.002

Dhanoa, M.A., Sanderson, R., Cardenas, L.M., Shepherd, A., Chadwick, D.R., Powell, C.D., Ellis, J.L., López, S., France, J., 2022. Chapter Five - Overview and application of the Mitscherlich equation and its extensions to estimate the soil nitrogen pool fraction associated with crop yield and nitrous oxide emission, in: Advances in Agronomy. Elsevier, pp. 269–295. https://doi.org/10.1016/bs.agron.2022.03.005

Diener, E., Lucas, R.E., Scollon, C.N., 2006. Beyond the hedonic treadmill: Revising the adaptation theory of well-being. Am. Psychol. 61, 305–314. https://doi.org/10.1037/0003-066X.61.4.305

Diener, E., Seligman, M.E.P., 2002. Very Happy People. Psychol. Sci. 13, 81–84. https://doi.org/10.1111/1467-9280.00415

Domingos, T., 2015. Accounting for carbon responsibility: the consumer and income perspectives and their reconciliation. Int. Input Output Assoc. Newsl. 32.

Doyal, L., Gough, I., 1991. A Theory of Human Need. Macmillan, Basingstoke.

Easterlin, R.A., 2015. Happiness and Economic Growth: The Evidence, in: Global Handbook of Quality of Life. Exploration of Well-Being of Nations and Continents, International Handbooks of Quality-of-Life. Springer Nature.

Easterlin, R.A., 1974. Does Economic Growth Improve the Human Lot? Some Empirical Evidence, in: Nations and Households in Economic Growth : Essays in Honor of Moses Abramovitz. Elsevier Science and Technology, pp. 89–125.

European Commission. Joint Research Centre., 2024. JRC statistical audit of the Commitment to reducing inequality index 2024. Publications Office, LU.

Farrell, M.J., 1957. The Measurement of Productive Efficiency. J. R. Stat. Soc. Ser. Gen. 120, 253. https://doi.org/10.2307/2343100

Fredrickson, B.L., 2001. The role of positive emotions in positive psychology: The broaden-and-build theory of positive emotions. Am. Psychol. 56, 218–226. https://doi.org/10.1037/0003-066X.56.3.218

Goutelle, S., Maurin, M., Rougier, F., Barbaut, X., Bourguignon, L., Ducher, M., Maire, P., 2008. The Hill equation: a review of its capabilities in pharmacological modelling. Fundam. Clin. Pharmacol. 22, 633–648. https://doi.org/10.1111/j.1472-8206.2008.00633.x

Harmacek, J., 2026. 2026 Global Social Progress Index. Social Progress Imperative, Washington, DC.

Hasannasab, M., Margaritis, D., Rouse, P., 2026. Benefit of the doubt models: some issues and insights. J. Product. Anal. 65, 14. https://doi.org/10.1007/s11123-026-00799-1

Helliwell, J.F., Layard, R., Sachs, J.D., De Neve, J.-E., Aknin, L.B., Wang, S. (Eds.), 2024. World Happiness Report 2024. Wellbeing Research Centre, University of Oxford, UK.

High-Level Expert Group on Beyond GDP, 2026. Counting What Counts A Compass of Progress for People and Planet. Report of the Secretary-General's independent High-Level Expert Group on Beyond GDP. United Nations, New York.

Hunter, L.E., Gershman, S.J., 2018. Reference-dependent preferences arise from structure learning. https://doi.org/10.1101/252692

Jansen, A., Wang, R., Behrens, P., Hoekstra, R., 2024. Beyond GDP: a review and conceptual framework for measuring sustainable and inclusive wellbeing. Lancet Planet. Health 8, e695–e705. https://doi.org/10.1016/S2542-5196(24)00147-5

Jebb, A.T., Tay, L., Diener, E., Oishi, S., 2018. Happiness, income satiation and turning points around the world. Nat. Hum. Behav. 2, 33–38. https://doi.org/10.1038/s41562-017-0277-0

Kahneman, D., Deaton, A., 2010. High income improves evaluation of life but not emotional well-being. Proc. Natl. Acad. Sci. 107, 16489–16493. https://doi.org/10.1073/pnas.1011492107

Keyes, C.L.M., 2002. The Mental Health Continuum: From Languishing to Flourishing in Life. J. Health Soc. Behav. 43, 207. https://doi.org/10.2307/3090197

Kraus, M., Tschernutter, D., Weinzierl, S., Zschech, P., 2024. Interpretable generalized additive neural networks. Eur. J. Oper. Res. 317, 303–316. https://doi.org/10.1016/j.ejor.2023.06.032

Kubiszewski, I., Costanza, R., Eastoe, J., Lu, T., Mulder, K., Hernandez, G.P., Benczûr, P., Dixson-Dectéve, S., 2025. Building consensus on societal wellbeing: a semantic synthesis of indicators to move beyond GDP. Ecol. Indic. 178, 114076. https://doi.org/10.1016/j.ecolind.2025.114076

Kubiszewski, I., Zakariyya, N., Costanza, R., Jarvis, D., 2020. Resilience of self-reported life satisfaction: A case study of who conforms to set-point theory in Australia. PLOS ONE 15, e0237161. https://doi.org/10.1371/journal.pone.0237161

Levy, D.J., Glimcher, P.W., 2012. The root of all value: a neural common currency for choice. Curr. Opin. Neurobiol. 22, 1027–1038. https://doi.org/10.1016/j.conb.2012.06.001

Lundberg, S., Lee, S.-I., 2017. A Unified Approach to Interpreting Model Predictions. https://doi.org/10.48550/ARXIV.1705.07874

Martínez-Alier, J., 1995. The environment as a luxury good or "too poor to be green"? Ecol. Econ. 13, 1–10. https://doi.org/10.1016/0921-8009(94)00062-Z

Mauro, V., Biggeri, M., Maggino, F., 2018. Measuring and Monitoring Poverty and Well-Being: A New Approach for the Synthesis of Multidimensionality. Soc. Indic. Res. 135, 75–89. https://doi.org/10.1007/s11205-016-1484-1

Max-Neef, M., 1991. Human-Scale Development: Conception, Application and Further Reflections. Apex, London.

Mazziotta, M., Pareto, A., 2019. Use and Misuse of PCA for Measuring Well-Being. Soc. Indic. Res. 142, 451–476. https://doi.org/10.1007/s11205-018-1933-0

Mazziotta, M., Pareto, A., 2016. On a Generalized Non-compensatory Composite Index for Measuring Socio-economic Phenomena. Soc. Indic. Res. 127, 983–1003. https://doi.org/10.1007/s11205-015-0998-2

McGregor, J.A., Pouw, N., 2016. Towards an economics of well-being. Camb. J. Econ. bew044. https://doi.org/10.1093/cje/bew044

Mizobuchi, H., 2014. Measuring World Better Life Frontier: A Composite Indicator for OECD Better Life Index. Soc. Indic. Res. 118, 987–1007. https://doi.org/10.1007/s11205-013-0457-x

Munda, G., 2008. Social Multi-Criteria Evaluation for a Sustainable Economy. Springer, Berlin, Heidelberg. https://doi.org/10.1007/978-3-540-73703-2

Munda, G., 2004. Social multi-criteria evaluation: Methodological foundations and operational consequences. Eur. J. Oper. Res. 158, 662–677. https://doi.org/10.1016/S0377-2217(03)00369-2

Munda, G., 1995. Multicriteria Evaluation in a Fuzzy Environment: Theory and Applications in Ecological Economics. Physica-Verlag HD.

Nardo, M., Saisana, M., Saltelli, A., Tarantola, S., Hoffmann, A., Giovannini, E., 2005. Handbook on Constructing Composite Indicators: Methodology and User Guide (OECD Statistics Working Papers), OECD Statistics Working Papers. https://doi.org/10.1787/533411815016

OECD, 2024. How's Life? 2024. Well-being and resilience in times of crisis. OECD Publishing, Paris.

OPHI, UNDP, 2025. Global Multidimensional Poverty Index 2025 Technical Note. United Nations development Programme and Ohford Poverty Development Institute.

Park, S.Q., Kahnt, T., Rieskamp, J., Heekeren, H.R., 2011. Neurobiology of Value Integration: When Value Impacts Valuation. J. Neurosci. 31, 9307–9314. https://doi.org/10.1523/JNEUROSCI.4973-10.2011

Pelt, D.H.M., Habets, P.C., Vinkers, C.H., Ligthart, L., Van Beijsterveldt, C.E.M., Pool, R., Bartels, M., 2024. Building machine learning prediction models for well-being using predictors from the exposome and genome in a population cohort. Nat. Ment. Health 2, 1217–1230. https://doi.org/10.1038/s44220-024-00294-2

Rockström, J., Steffen, W., Noone, K., Persson, Å., Chapin, F.S., Lambin, E.F., Lenton, T.M., Scheffer, M., Folke, C., Schellnhuber, H.J., Nykvist, B., de Wit, C.A., Hughes, T., van der Leeuw, S., Rodhe, H., Sörlin, S., Snyder, P.K., Costanza, R., Svedin, U., Falkenmark, M., Karlberg, L., Corell, R.W., Fabry, V.J., Hansen, J., Walker, B., Liverman, D., Richardson, K., Crutzen, P., Foley, J.A., 2009. A safe operating space for humanity. Nature 461, 472–475. https://doi.org/10.1038/461472a

Roy, B., 1968. Classement et choix en présence de points de vue multiples: La méthode ELECTRE. Rev. Francaise Inform. Rech. Opérationnelle 8, 57–75.

Rutledge, R.B., Skandali, N., Dayan, P., Dolan, R.J., 2014. A computational and neural model of momentary subjective well-being. Proc. Natl. Acad. Sci. 111, 12252–12257. https://doi.org/10.1073/pnas.1407535111

Ryan, R.M., Deci, E.L., 2001. On Happiness and Human Potentials: A Review of Research on Hedonic and Eudaimonic Well-Being. Annu. Rev. Psychol. 52, 141–166. https://doi.org/10.1146/annurev.psych.52.1.141

Ryff, C.D., 2014. Self-realisation and meaning making in the face of adversity: a eudaimonic approach to human resilience. J. Psychol. Afr. 24, 1–12. https://doi.org/10.1080/14330237.2014.904098

Ryff, C.D., Keyes, C.L.M., 1995. The structure of psychological well-being revisited. J. Pers. Soc. Psychol. 69, 719–727. https://doi.org/10.1037/0022-3514.69.4.719

Saisana, M., Saltelli, A., Tarantola, S., 2005. Uncertainty and Sensitivity Analysis Techniques as Tools for the Quality Assessment of Composite Indicators. J. R. Stat. Soc. Ser. A Stat. Soc. 168, 307–323. https://doi.org/10.1111/j.1467-985X.2005.00350.x

Sands, D.C., Morris, C.E., Dratz, E.A., Pilgeram, A.L., 2009. Elevating optimal human nutrition to a central goal of plant breeding and production of plant-based foods. Plant Sci. 177, 377–389. https://doi.org/10.1016/j.plantsci.2009.07.011

Seghouani, M., Bravin, M.N., Mollier, A., 2024. Crop response to nitrogen-phosphorus colimitation: theory, experimental evidences, mechanisms, and models. A review. Agron. Sustain. Dev. 44, 3. https://doi.org/10.1007/s13593-023-00939-z

Seligman, M.E.P., 2011. Flourish: a visionary new understanding of happiness and well-being, 1. Free Press hardcover ed. ed. Free Press, New York, NY.

Seligman, M.E.P., Csikszentmihalyi, M., 2000. Positive psychology: An introduction. Am. Psychol. 55, 5–14. https://doi.org/10.1037/0003-066X.55.1.5

Steffen, W., Richardson, K., Rockström, J., Cornell, S.E., Fetzer, I., Bennett, E.M., Biggs, R., Carpenter, S.R., de Vries, W., de Wit, C.A., Folke, C., Gerten, D., Heinke, J., Mace, G.M., Persson, L.M., Ramanathan, V., Reyers, B., Sörlin, S., 2015. Planetary boundaries: Guiding human development on a changing planet. Science 347, 1259855. https://doi.org/10.1126/science.1259855

Steinberger, J.K., Roberts, J.T., 2010. From constraint to sufficiency: The decoupling of energy and carbon from human needs, 1975-2005. Ecol. Econ. 70, 425–433. https://doi.org/10.1016/j.ecolecon.2010.09.014

Stiglitz, J.E., Sen, A., Fitoussi, J.-P., 2009. Report by the commission on the measurement of economic performance and social progress. Commission on the Measurement of Economic Performance and Social Progress.

Stirling, A., 2006. Analysis, participation and power: justification and closure in participatory multi-criteria analysis. Land Use Policy 23, 95–107. https://doi.org/10.1016/j.landusepol.2004.08.010

United Nations Development Programme, 2026. Planetary pressures–adjusted Human Development Index (PHDI). UNDP.

Walheer, B., 2024. A sequential benefit-of-the-doubt composite indicator. Eur. J. Oper. Res. 316, 228–239. https://doi.org/10.1016/j.ejor.2024.01.029

You, S., Ding, D., Canini, K., Pfeifer, J., Gupta, M., 2017. Deep Lattice Networks and Partial Monotonic Functions. https://doi.org/10.48550/ARXIV.1709.06680

**Supplementary Table S1 - Minimum, median, and maximum normalised scored for SIW components**

| Countries | $CO_2$_min | $CO_2$_med | $CO_2$_mean | $CO_2$_max | MF_min | MF_med | MF_mean | MF_max | LE_min | LE_med | LE_mean | LE_max | EDU_min | EDU_med | EDU_mean | EDU_max | INC_min | INC_med | INC_mean | INC_max |
|---|---|---|---|---|---|---|---|---|---|---|---|---|---|---|---|---|---|---|---|---|
| A | 0.75 | 0.68 | 0.68 | 0.74 | 0.71 | 0.60 | 0.60 | 0.63 | 1.00 | 1.00 | 1.00 | 1.00 | 1.00 | 1.00 | 1.00 | 1.00 | 1.00 | 1.00 | 1.00 | 0.74 |
| B | 0.90 | 0.88 | 0.87 | 0.90 | 0.88 | 0.85 | 0.84 | 0.86 | 0.90 | 0.89 | 0.89 | 0.93 | 0.89 | 0.90 | 0.90 | 0.87 | 0.53 | 0.65 | 0.65 | 0.41 |
| C | 0.00 | 0.00 | 0.00 | 0.00 | 0.00 | 0.00 | 0.00 | 0.00 | 0.71 | 0.74 | 0.74 | 0.86 | 0.67 | 0.71 | 0.70 | 0.78 | 0.77 | 0.90 | 0.93 | 1.00 |
| D | 0.53 | 0.53 | 0.53 | 0.53 | 0.59 | 0.47 | 0.48 | 0.49 | 0.81 | 0.84 | 0.84 | 0.86 | 0.78 | 0.80 | 0.80 | 0.78 | 0.48 | 0.55 | 0.54 | 0.38 |
| E | 0.98 | 0.97 | 0.97 | 0.97 | 0.94 | 0.94 | 0.94 | 0.96 | 0.86 | 0.95 | 0.95 | 1.00 | 0.96 | 0.95 | 0.98 | 0.96 | 0.42 | 0.45 | 0.43 | 0.31 |
| F | 0.96 | 0.93 | 0.93 | 0.94 | 0.82 | 0.81 | 0.80 | 0.82 | 0.95 | 0.95 | 0.95 | 0.93 | 0.91 | 0.90 | 0.93 | 0.87 | 0.30 | 0.35 | 0.34 | 0.21 |
| G | 0.84 | 0.77 | 0.77 | 0.77 | 0.76 | 0.68 | 0.68 | 0.69 | 0.62 | 0.63 | 0.63 | 0.71 | 0.56 | 0.61 | 0.63 | 0.65 | 0.13 | 0.20 | 0.21 | 0.18 |
| H | 0.88 | 0.83 | 0.72 | 0.53 | 0.82 | 0.77 | 0.64 | 0.29 | 0.24 | 0.53 | 0.47 | 0.71 | 0.22 | 0.49 | 0.45 | 0.78 | 0.04 | 0.10 | 0.27 | 0.54 |
| I | 1.00 | 1.00 | 1.00 | 1.00 | 1.00 | 1.00 | 1.00 | 1.00 | 0.00 | 0.00 | 0.00 | 0.00 | 0.00 | 0.00 | 0.00 | 0.00 | 0.00 | 0.00 | 0.00 | 0.00 |
| J | 0.80 | 0.71 | 0.66 | 0.58 | 0.65 | 0.55 | 0.52 | 0.39 | 0.33 | 0.42 | 0.42 | 0.57 | 0.33 | 0.41 | 0.40 | 0.52 | 0.10 | 0.15 | 0.19 | 0.28 |

Abbreviations: $CO_2$ – $CO_2$ emissions, MF – Material flows, LE – Life expectancy, EDU – Education, INC – Income, min – minimum value, med – median value, max – maximum value.